\documentclass[conference]{IEEEtran}
\IEEEoverridecommandlockouts

\usepackage{tikz,pgfplots}
\pgfplotsset{compat=newest}
\usepackage{amsmath,amsthm}
\usepackage{amssymb}
\usepackage{wasysym}
\usepackage{rotating}
\usepackage[colorinlistoftodos]{todonotes}
\definecolor{darkgreen}{rgb}{0.12549019607843137255,0.4980392156862745098,0.16862745098039215686}
\usetikzlibrary{arrows,calc,shadows,positioning,shapes.geometric,decorations.pathmorphing,decorations.shapes,decorations.markings,patterns}

\usepackage{fixltx2e}

\ifCLASSINFOpdf
\else
\fi
\begin{document}
%
\title{Compressed Sensing Bayes Risk Minimization for Under-determined Systems via Sphere Detection}


\author{\IEEEauthorblockN{Fabian Monsees, Carsten Bockelmann, Dirk W\"{u}bben, Armin Dekorsy}\\
\IEEEauthorblockA{Department of Communications Engineering\\
University of Bremen, Germany\\
Email: \{monsees, bockelmann, wuebben, dekorsy\}@ant.uni-bremen.de\\
}}


%


\maketitle

\begin{abstract}
The application of Compresses Sensing is a promising physical layer technology for the joint activity and data detection of signals. Detecting the activity pattern correctly has severe impact on the system performance and is therefore of major concern. In contrast to previous work, in this paper we optimize joint activity and data detection in under-determined systems by minimizing the Bayes-Risk for erroneous activity detection. We formulate a new Compressed Sensing Bayes-Risk detector which directly allows to influence error rates at the activity detection dynamically by a parameter that can be controlled at higher layers. We derive the detector for a general linear system and show that our detector outperforms classical Compressed Sensing approaches by investigating an overloaded CDMA system.

\end{abstract}


%
\IEEEpeerreviewmaketitle

\section{Introduction}
Compressed Sensing (CS) is seen as a promising technology towards novel detectors, optimized for the detection of sparse signals even in under determined systems. Various CS algorithms formulate the problem of solving an under-determined set of equations as a convex optimization problem~\cite{yonia}. Involving a sparsity promoting term such as the $l_1$ norm guarantees that the algorithm converges and finds the sparsest solution to an under-determined system~\cite{4472240}. Besides that, various greedy approaches exist that solve under-determined problems iteratively. The most prominent ones are the Orthogonal Matching Pursuit (OMP)~\cite{342465} and the Orthogonal Least Squares (OLS)~\cite{eps251147}.
\\
In the field of communication technology, e.g, CS was successfully applied to OFDM systems for the reduction of the Peak to Average Power Ratio~\cite{6292852}. Furthermore, in~\cite{henn1} the authors have shown how novel CS techniques can be applied to Machine-to-Machine communication setups in order to detect activity and data jointly, thereby decreasing the required signaling overhead. 
\\
Joint detection of activity and data opens a wide field of application for CS algorithms in communications. Detecting sparse multi-user signals in a communication system is mostly motivated by assuming that some users are inactive and, thus, modeled as transmitting zeros, making the signal sparse. However, in such a scenario, the detection task is beyond finding the sparsest solution as pursued in CS. Rather, the detector has to estimate which users are active and which are inactive. The detection of the correct activity pattern, known as vector support in CS, is therefore crucial for successful Compressed-Sensing Multi-User Detection (CS-MUD). Another important point is that signals in communication obey a modulation alphabet which is generally a non-convex set. Accordingly, problems in communication have to be relaxed prior to solving them by common CS solvers. 
\\
With this application in mind, we would like to stress that wrongly detecting a user to be inactive is generally worse than doing the opposite, i.e., wrongly detecting a user to be active. In the first case information is lost, whereas higher layer error processing can help to identify erroneous transmissions in the latter case. The so called False Active and False Inactive rate is consequently of crucial importance for CS-MUD.
\\
In this paper, we derive a Bayes-Risk detector for under-determined systems with non-convex finite alphabet constraints, which allows us to control the performance of the activity detection with respect to False Active and False Inactive errors. Moreover, this Bayes-Risk CS-MUD (BR-CS-MUD) can efficiently be implemented using common tree search algorithms such as the Sphere-Detector. As an example, we show the performance of our BR-CS-MUD detector by investigating an overloaded Code Division Multiple Access (CDMA) network, which can be expressed as an under-determined set of equations. The proposed detector utilizes the information about a finite modulation alphabet and finds the optimal solution by finding the closest lattice point. Utilizing this strategy outperforms common convex CS solvers such as the Basis Pursuit De-noising (BPDN) approach.

\section{Compressed Sensing Bayes-Risk Detection}
\subsection{Sporadic Communication Model}
In the following we shortly outline the sporadic communication model which forms the bases for the formulation of the Bayes-Risk Compressed-Sensing detector.
In this model we aim to recover an unknown source vector $\mathbf{x}$ whose elements obey an augmented finite modulation alphabet $\mathcal{A}_0 = \mathcal{A} \cup \{0\}$. Here $\mathcal{A}$ stands for the modulation alphabet that is used for data transmission and can be, e.g., Binary Phase Shift Keying. The zero is used to model inactivity in the source vector which we assume to be of dimension $\mathbf{x} \in \mathcal{A}_0^K$. In a communication context, $\mathbf{x}$ can contain the symbols from different nodes in a multi-user scenario~\cite{henn1} or symbols from different antennas in a multiple antenna context. We further assume that the elements $x_k$ of $\mathbf{x}$ are i.i.d and each element is non-zero with probability $p_a$. Here $p_a$ can be interpreted as an activity probability. With low $p_a$, many elements of $\mathbf{x}$ contain zeros, which makes the vector sparse. 
We summarize the channels between the nodes and the aggregation node with the matrix $\mathbf{T}\in \mathbb{R}^{M \times K}$.  $\mathbf{y} \in \mathbb{R}^M$  represents the noise corrupted observation at the aggregation node. The canonical input-output relation of the system is given by

\begin{equation}\label{eq:general_syste_model}
 \mathbf{y}=\mathbf{Tx}+\mathbf{w}.
\end{equation}
If $M < K$ the system refers to an under-determined  system where the number of variables exceeds the number of observations.
The noise vector $\mathbf{w}\! \sim \! \mathcal{N}\left(\sigma_n^2,0\right)$ obeys an uncorrelated zero mean Gaussian distribution with variance $\sigma_n^2$. Henceforth, the task is to recover $\mathbf{x} \in \mathcal{A}_0^K$ from $M$ noise corrupted measurements. To ease notation, we consider the communication system to be a real valued system which is not a general restriction. The scheme we propose can also be applied to complex systems using an equivalent real valued system description. With~\eqref{eq:general_syste_model} we have a common noisy CS problem with a-priori information about the source vector. 

\subsection{Generalized Likelihood Ratio Test}
With the previously introduced sporadic communication model, the goal of the detector is to reconstruct the symbols from the augmented modulation alphabet $\mathcal{A}_0$. In~\cite{giannakis} the authors derived the MAP detector for an overdetermined system given the knowledge about the a priori distribution of the source vector $\mathbf{x}$. Our approach to recover $\mathbf{x}$ is twofold. First, we derive a detector that copes with under-determined systems given the a-priori knowledge of the source vector. Second, we formulate the detector such that the error rates for the activity detection can be tuned depending on the system by formulating the detection task in a Bayesian framework.
\\
We start by considering the activity detection. For each element $\hat x_k$ the detector estimates whether the element was active, i.e., $\hat x_k \in \mathcal{A}$ or inactive, i.e., $\hat x_k=0$. The possible outcomes for the activity detection task are summarized as follows.
\begin{table}[h!]
\begin{center}
\begin{tabular}{r|c|c|}
\multicolumn{1}{r}{}
 &  \multicolumn{1}{c}{$x_k \in \mathcal{A}$}
 & \multicolumn{1}{c}{$x_k = 0$} \\
\cline{2-3}
$\hat x_k \in \mathcal{A}$ & True Active & False Active \\
\cline{2-3}
$\hat x_k = 0$ & False Inactive & True Inactive \\
\cline{2-3}
\end{tabular}
\caption{Confusion Matrix for Activity Detection}\label{table_conf_matrix}
\end{center}
\vspace{-3.5em}
\end{table}
Table~\ref{table_conf_matrix} shows the so called Confusion matrix of the possible mappings between the true classes (element is active or inactive) and the hypothesized classes (element is active or inactive)~\cite{citeulike:161814}. The two error events are of special interest. If an element is wrongly estimated to be inactive, the data corresponding to this element is simply lost and cannot be recovered. However, in the the opposite case, when an element is wrongly estimated to be active, higher layer error detection methods can sort out wrong data packets. Thus, we state that these two possible error events have different impact on the overall system performance. With the knowledge that the two error classes have different impact, our approach is to derive a detector that can be controlled by an additional parameter that balances the activity detection.
\\
To include this balancing in the detector, we consider a binary hypothesis testing problem arising when determining whether an element is active or inactive.
\begin{align}
 & H1: x_k  \in \mathcal{A} \rightarrow \text{Element is active} \notag\\
 & H2: x_k = 0 \rightarrow \text{Element is inactive} \notag
\end{align}
With Tab.~\ref{table_conf_matrix} we assign the following costs for wrong activity detection.
\begin{itemize}
 \item $C_{\text{Fi}}$ : Cost for estimating False Inactive
 \item $C_{\text{Fa}}$ : Cost for estimating False Active
\end{itemize}
Costs for correct decisions are implicitly set to zero. We cast the problem as the minimization of the Bayes-Risk defined as
\begin{align}
 \mathcal{R} &= C_{\text{Fa}}\text{Pr}\left(x_k=0\right)\int_{Z_{\mathcal{A}}}p\left(y_m | x_k = 0\right)dy_m + \notag\\
&C_{\text{Fi}}\text{Pr}\left(x_k \in \mathcal{A}\right)\int_{Z_0}p\left(y_m | x_k \in \mathcal{A}\right)dy_m. \label{eq:bayes_risk}
\end{align}
The Bayes-Risk given in~\eqref{eq:bayes_risk} expresses the risk of a wrong activity detection. $p\left(y_m|x_k\right)$ denotes the density for an observation $y_m$ under hypothesis $x_k$. $\text{Pr}\left(\cdot\right)$ is an event probability. $Z_0$ and $Z_\mathcal{A}$ are the regions in the observation space where the detector estimates the element to be inactive $\hat x_k=0$ or active $\hat x_k \in \mathcal{A}$. For the sake of clarity we omit lengthy derivations and refer the reader to~\cite{VanTDeteRada1971}. 
With the results from~\cite{stephem_kay} we cast~\eqref{eq:bayes_risk} as a Generalized Likelihood Ratio Test (GLRT) which has the following form
\begin{equation}\label{eq:GLRT}
 \frac{p\left(y_m|x_k=0\right)\text{Pr}\left(x_k=0\right)C_{\text{Fa}}}{\max_{x\in \mathcal{A}} p\left(y_m|x_k\right)\text{Pr}\left(x_k\right)C_{\text{Fi}}} \quad\mathop{\gtrless}_{H1}^{H2} 1 
\end{equation}
The GLRT in~\eqref{eq:GLRT} evaluates the weighted a posteriori probabilities for both hypothesis and  assigns the observation to the hypothesis with higher probability. Moreover, hypothesis $H1$ covers the set $\mathcal{A}$ and is therefore a so-called composite hypothesis. As~\eqref{eq:GLRT} shows we compare the best model in $H1$ with $H2$. It can be seen that~\eqref{eq:GLRT} can be rewritten as
\begin{equation}\label{eq:elementwise MAP}
 \hat x_k = \underset{x_k \in \mathcal{A}_0}{\text{arg max}}\ p\left(y_m | x_k\right)\text{Pr}\left(x_k\right)C\left(x_k\right),
\end{equation}
with 
\begin{equation}\label{eq_cost_function}
 C\left(x_k\right)= C_{\text{Fi}}^{\mathbf{1}_{\mathcal{A}}\left(x_k\right)}C_{\text{Fa}}^{1-\mathbf{1}_{\mathcal{A}}\left(x_k \right)}.
\end{equation}
Here, $\mathbf{1}_{\mathcal{A}}\left(\cdot\right)$ is the indicator function which takes the value $1$ if the argument is contained in the set $\mathcal{A}$. The solution to the optimization problem given in~\eqref{eq:elementwise MAP} is the optimal scalar estimate for $x_k$ given a scalar observation $y_m$. Assuming i.i.d. observations and vector hypothesis $\mathbf{x}$, it can be shown that under application of the max-log approximation,~\eqref{eq:elementwise MAP} can be extended to
\begin{equation}\label{eq:optimization:task}
   \mathbf{\hat x} = \underset{\mathbf{x} \in \mathcal{A}_0}{\text{arg max}}\ \log \prod_{m=1}^{M}p\left(y_m | \mathbf{x}\right)\prod_{k=1}^{K}\text{Pr}\left(x_k\right)C\left(x_k\right).
\end{equation}
In the following we consider the system given in~\eqref{eq:general_syste_model}. The a posteriori probability for the element $y_m$  with vector hypothesis $\mathbf{x}$ can now be expressed as
\begin{equation}
 p\left(y_m|\mathbf{x}\right) = \frac{1}{\sqrt{2\pi}\sigma_n}\exp\left[-\frac{1}{2\sigma_n^2} | y_m - \mathbf{T}_m\mathbf{x} |^2 \right],
\end{equation}
and
\begin{equation}\label{eq:gaussian_posteri}
 \prod_{m=1}^{M} p\left(y_m|\mathbf{x}\right) = \frac{1}{(2\pi)^{M/2}\sigma_n^{M}}\exp\left[-\frac{1}{2\sigma_n^2}\|\mathbf{y}-\mathbf{Tx}\|_2^2\right].
\end{equation}
Here $\mathbf{T}_m$ is the $m$th row vector of the matrix $\mathbf{T}$. 
In the following we assume that the elements of the source vector $x_k$ from~\eqref{eq:general_syste_model} obey a Bernoulli / Uniform distribution. In particular, we have
\begin{align}
&\text{Pr}\left(x_k \in \mathcal{A}\right) =1-\text{Pr}\left(x_k=0\right)=p_a \\
&p\left(x_k |x_k \in \mathcal{A}\right) =\frac{1}{|\mathcal{A}|} 
\end{align}
and we write the a priori probability of the source vector $\mathbf{x}$ weighted with the costs from~\eqref{eq_cost_function} as
\begin{align}
 &\prod_{k=1}^{K}\text{Pr}\left(x_k\right)C\left(x_k\right) =\notag\\
 &\prod_{k=1}^{K}\left[C_{\text{Fa}}\left(1-p_a\right)\right]^{1-\mathbf{1}_{\mathcal{A}}\left(x_k \right)}\left(C_{\text{Fi}}\frac{p_a}{|\mathcal{A}|}\right)^{\mathbf{1}_{\mathcal{A}}\left(x_k\right)}\notag\\
&=\left[C_{\text{Fa}}\left(1-p_a\right)\right]^{K-\sum_{k}\mathbf{1}_{\mathcal{A}}\left(x_k\right)} \left(C_{\text{Fi}}\frac{p_a}{|\mathcal{A}|}\right)^{\sum_{k}\mathbf{1}_{\mathcal{A}}\left(x_k\right)},\label{eq:prior_on_x}
\end{align}
with~\eqref{eq:gaussian_posteri} and~\eqref{eq:prior_on_x} we can finally rewrite the optimization problem given in~\eqref{eq:optimization:task} to 
\begin{equation}
 \mathbf{\hat x} = \underset{\mathbf{x} \in \mathcal{A}_0}{\text{arg min}} \|\mathbf{y}-\mathbf{Tx}\|_2^2 + 2\sigma_n^2 \sum_{k=1}^{K}\mathbf{1}_{\mathcal{A}}\left(x_k\right)\log\left(\frac{C_{\text{Fa}}}{C_{\text{Fi}}}\frac{1-p_a}{p_a/{|\mathcal{A}|}}\right).\label{eq:map_with_indexing function}
\end{equation}
Moreover, the indicator function $\mathbf{1}_{\mathcal{A}}\left(x_k\right)$ can be replaced by the zero pseudo-norm defined as
\mbox{$\|\mathbf{x}\|_0 = \#\{x_k : x_k \ne 0\}$}. We further express the ratio of costs as $\Omega=\frac{C_{\text{Fa}}}{C_{\text{Fi}}}$ and we obtain the non-convex optimization problem 
\begin{align}\label{eq:map:withl0:norm}
 \mathbf{\hat x} &= \underset{\mathbf{x} \in \mathcal{A}_0}{\text{arg min}} \|\mathbf{y}- \mathbf{T}\mathbf{x}\|_2^2 + 2\sigma_n^2\|\mathbf{x}\|_0\text{log}\left(\Omega\frac{1-p_a}{p_a/{|\mathcal{A}|}}\right)\notag\\
 \mathbf{\hat x} &= \underset{\mathbf{x} \in \mathcal{A}_0}{\text{arg min}} \|\mathbf{y}- \mathbf{T}\mathbf{x}\|_2^2 + \lambda\left(\Omega\right)\|\mathbf{x}\|_0.
\end{align}
Where the penalty term $\lambda\left(\Omega\right)=2\sigma_n^2\text{log}\left(\Omega\frac{1-p_a}{p_a/{|\mathcal{A}|}}\right)$ reflects the a priori assumption about the source vector $\mathbf{x}$ weighted by the parameter $\Omega$. With $\Omega=1$ and $M \ge K$~\eqref{eq:map:withl0:norm} equals to the MAP detector from~\cite{giannakis}. The optimization problem~\eqref{eq:map:withl0:norm} can be interpreted as a generalized MAP detector involving the Bayes-Risk for the activity detection. Additionally, $\Omega$ scales the a priori assumption and can be interpreted as an additional parameter that determines whether the detector is conservative $\Omega > 1$ or liberal $\Omega < 1$. A conservative detector will decide in favor of inactivity and will produce less False Active errors. In contrast a liberal detector will decide in favor of activity and produce less False Inactive errors than a conservative detector. The problem given in~\eqref{eq:map:withl0:norm} is non-convex but can be implemented via a Sphere-
Detector under the condition that $M \ge K$ holds. In the following we augment this detector for under-determined systems, i.e., $M < K$.

\subsection{Constant Modulus Restriction}
To solve~\eqref{eq:map:withl0:norm} for $M<K$ efficiently, we restrict to constant modulus alphabets which allows us to replace the $l_0$ pseudo-norm by the $l_p$ norm, i.e.,
$\|\mathbf{x}\|_0 =\|x\|_2^2= \|x\|_p^p$
and we have
\begin{equation}\label{eq:map_with_l2}
 \mathbf{\hat x} = \underset{\mathbf{x} \in \mathcal{A}_0}{\text{arg min}} \|\mathbf{y}- \mathbf{T}\mathbf{x}\|_2^2 + \lambda\left(\Omega\right)\|\mathbf{x}\|_p^p.
\end{equation}
The penalty term $\lambda(\Omega)$ is directly influenced by $\Omega$ and can be negative for small values of $\Omega$. We make use of the norm-invariance of the source vector $\mathbf{x}$ and expand~\eqref{eq:map_with_l2} to an overdetermined system 
\begin{align}
 \mathbf{\hat x} &= \underset{\mathbf{x} \in \mathcal{A}_0}{\text{arg min}} \|\mathbf{y}- \mathbf{T}\mathbf{x}\|_2^2 + \|\mathbf{x}\|_2^2+ \left[\lambda\left(\Omega\right)-1\right]\|\mathbf{x}\|_0\\
&= \left\Vert{\left[\hspace{-0.5em}\mathbf{y} \atop \mathbf{0}_K\right]} - {\left[\hspace{-0.5em}\mathbf{T} \atop \mathbf{I}_K\right]}\mathbf{x}\right\Vert_2^2+ \underbrace{\left[\lambda\left(\Omega\right)-1\right]}_{\Theta\left(\Omega\right)}\|\mathbf{x}\|_0. \label{eq:regularized:detector}
\end{align}
If $\Theta(\Omega) \ge 0$,~\eqref{eq:regularized:detector} can be directly implemented using a Sphere-Detector. One of the main prerequisites for the Sphere-Detector is an increasing metric in the sequence of detected symbols which is not guaranteed if $\Theta(\Omega) <0$. However, $\Theta(\Omega)$ controls how non-zero symbols are penalized compared to no costs for zero symbols. For $\Theta(\Omega) <0$ the cost for non-zero symbols are set to a negative value while zero symbols are still charged with no costs. To ensure that the penalty term remains positive, we re-write the optimization problem such that zero-symbols are penalized with positive costs. To proof this approach, we briefly review that the $l_0$ pseudo norm can be written as a sequence of indicator functions as $\|\mathbf{x}\|_0 = \sum_{k=1}^K\mathbf{1}_{\mathcal{A}}\left(x_k\right)$. If $\Theta(\Omega)<0$, we can rewrite the optimization problem by subtracting the absolute value of the negative costs while adding the constant $|\Theta(\Omega)
|K$. These two step do not change the optimization problem and we obtain for~\eqref{eq:regularized:detector}
\begin{align}
  \mathbf{\hat x} &= \underset{\mathbf{x} \in \mathcal{A}_0}{\text{arg min}} \left\Vert{\left[\hspace{-0.5em}\mathbf{y} \atop \mathbf{0}_K\right]}- {\left[\hspace{-0.5em}\mathbf{T} \atop \mathbf{I}_K\right]}\mathbf{x}\right\Vert_2^2   -|\Theta\left(\Omega\right)| \|\mathbf{x}\|_0 +|\Theta(\Omega)|K\notag\\
 &= \underset{\mathbf{x} \in \mathcal{A}_0}{\text{arg min}} \left\Vert{\left[\hspace{-0.5em}\mathbf{y} \atop \mathbf{0}_K\right]}- {\left[\hspace{-0.5em}\mathbf{T} \atop \mathbf{I}_K\right]}\mathbf{x}\right\Vert_2^2+  |\Theta\left(\Omega\right)|\left[K- \|\mathbf{x}\|_0 \right] \notag\\
  &=\underset{\mathbf{x} \in \mathcal{A}_0}{\text{arg min}} \left\Vert{\left[\hspace{-0.5em}\mathbf{y} \atop \mathbf{0}_K\right]}- {\left[\hspace{-0.5em}\mathbf{T} \atop \mathbf{I}_K\right]}\mathbf{x}\right\Vert_2^2+  |\Theta\left(\Omega\right)| \sum_{k=1}^K\mathbf{1}_{0}\left(x_k\right)\notag\\
&=\underset{\mathbf{x} \in \mathcal{A}_0}{\text{arg min}} \|\mathbf{y}^{\prime}- \mathbf{T}^{\prime} \mathbf{x}\|_2^2+  |\Theta\left(\Omega\right)| \sum_{k=1}^K\mathbf{1}_{0}\left(x_k\right).\label{eq:map_theta<0}
\end{align}
Due to the application of $\mathbf{1}_{0}\left(\cdot\right)$, each zero symbol is penalized by $\Theta(\Omega)$. Moreover,  the penalty term in~\eqref{eq:map_theta<0} shows that the metric is monotonically increasing in the elements contained in $\mathbf{x}$ which allows direct implementation via a Sphere-Detector. We note that the augmented matrices $\mathbf{y}^{\prime}\in \mathbb{R}^{M+K}$ and $\mathbf{T}^{\prime} \in \mathbb{R}^{M+K \times K}$ constitute an overdetermined set of equations.

\subsection{Compressed-Sensing Bayes-Risk Sphere-Detector}
We here briefly rewrite the optimization problem~\eqref{eq:map_theta<0} such that efficient implementation via a Sphere-Detector is possible. We apply the following procedure for solving~\eqref{eq:map_theta<0} efficiently. First, we apply the skinny \mbox{\cite{Golub1989}[p. 217]} QR decomposition on $\mathbf{T}^{\prime}$ and obtain $\mathbf{T}^{\prime}=\mathbf{QR}$ with $\mathbf{Q} \in \mathbb{R}^{M+K \times K}$ being a matrix with orthonormal columns and $\mathbf{R} \in \mathbb{R}^{K \times K}$ being an upper triangular matrix. 
With this decomposition of $\mathbf{T}^{\prime}$, we rewrite the optimization problem~\eqref{eq:map_theta<0} as 
\begin{align}\label{eq:map:QR}
&=\underset{\mathbf{x} \in \mathcal{A}_0}{\text{arg min}} \|\mathbf{y}^{\prime}- \mathbf{QR}\mathbf{x}\|_2^2+  |\Theta\left(\Omega\right)| \sum_{k=1}^K\mathbf{1}_{0}\left(x_k\right)\\
&=\underset{\mathbf{x} \in \mathcal{A}_0}{\text{arg min}} \|\mathbf{Q}^T\mathbf{y}^{\prime}- \mathbf{Q}^T\mathbf{QR}\mathbf{x}\|_2^2+  |\Theta\left(\Omega\right)| \sum_{k=1}^K\mathbf{1}_{0}\left(x_k\right)\\
&=\underset{\mathbf{x} \in \mathcal{A}_0}{\text{arg min}} \|\mathbf{\tilde y}- \mathbf{R}\mathbf{x}\|_2^2+  |\Theta\left(\Omega\right)| \sum_{k=1}^K\mathbf{1}_{0}\left(x_k\right)
\end{align}
Note that for a $\Theta(\Omega)\ge 0$, the optimization problem can be solved by the direct application of the QR decomposition on ~\eqref{eq:regularized:detector}~\cite{giannakis}\cite{damen}.
\section{Performance Evaluation}
\vspace{-1em}
\subsection{Setup}
In the following, we exemplary show the performance of our detector by investigating an overloaded Code Division Multiple Access (CDMA)~\cite{Verdu} system. In this model, we assume that in total $K$ sensor nodes are connected to a central aggregation node. The nodes are  active only occasionally with a probability factor $p_a$ which is equal for all $K$ nodes in the system. Active nodes transmit Binary Phase Shift Keying (BPSK) modulated symbols and we have $\mathcal{A}=\{\pm 1\}$. The symbols are spread by pseudo-noise sequences to chips by a factor of $N$ which is equal for all nodes in the network. The chips are transmitted over a frequency selective Rayleigh fading channel with impulse response length of $L_h$ which is also assumed to be of equal length for all nodes. Inactive nodes are modeled as transmitting zeros. The joint activity and data detection task is thus the detection of symbols from the augmented alphabet $\mathcal{A}_0=\mathcal{A} \cup \{0\}$. In the following we write $\mathbf{x} \in \
mathcal{A}_0^K$ as the source vector containing the symbols from the $K$ nodes. To be consistent with the previously derived detector, we express the node specific spreading and convolution with the underlaying channel with the matrix $\mathbf{T} \in \mathbb{R}^{N+L_h-1 \times K}$.
This allows us to formulate the symbol-rate CDMA system as
\begin{equation}
\mathbf{y} = \mathbf{T}\mathbf{x} + \mathbf{ w}.\label{eq:system:model:after:prewh}
\end{equation}
Here $\mathbf{y} \in \mathbb{R}^{N + L_h -1}$ is the received signal at chip-rate and $\mathbf{ w}$ is the white uncorrelated noise vector with zero mean and variance $\sigma_n^2$. We apply a pre-whitening filter in order to ensure white Gaussian noise at symbol-rate. We neglect the details here, for further information the reader is referred to~\cite{mons1}.

\subsection{Symbol Error Rate}
We investigate the performance of the detector by considering errors on the augmented alphabet $\mathcal{A}_0$ which we term Gross Symbol Errors (GSE) since activity and data errors are summarized in this class. For assessing the performance at the GSE, we start by investigating it for different degrees of overloading the system. Overloading is done by varying $N$ while keeping $K$ constant. The remaining simulation parameters are summarized in Table~\ref{table_sim_par1}
\begin{table}[ht]
  \begin{tabular}{| c | c |}
  \hline
  \multicolumn{2}{|c|}{Simulation Parameters} \\
    \hline
    \hline
    Number of Nodes  & $K=20$  \\    \hline
    Spreading Gain & $1 \le N \le 20$  \\    \hline
    Length of Channel Impulse Resp. & $L_h=4$ chips  \\   \hline
    Channel Type & real valued Rayleigh Fading   \\ \hline
    Channel State Information & Perfect   \\ \hline
    Activity Probability & $p_a=0.2$  \\   \hline
    Bayes Factor & $\Omega =1$  \\   \hline
    Modulation Type & BPSK  \\ 
    \hline
  \end{tabular}
\caption{Simulation Parameter}\label{table_sim_par1}
\vspace{-2.5em}
\end{table}

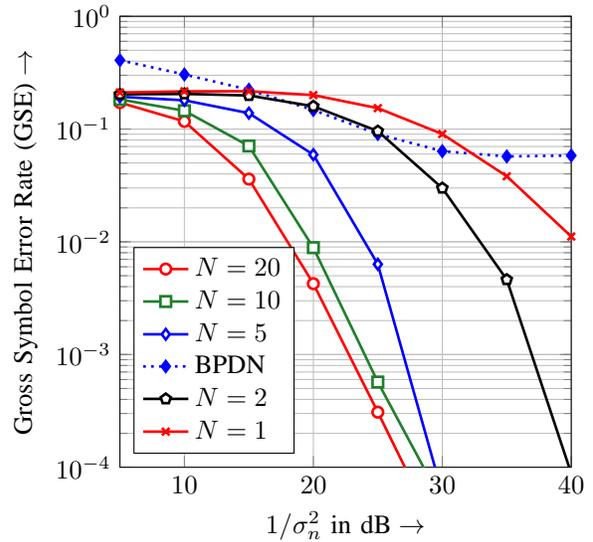
\begin{figure}[ht]
%
%
\begin{tikzpicture}

\begin{semilogyaxis}[%
scale only axis,
width=6cm,
height=6cm,
xmin=5, xmax=40,
ymin=0.0001, ymax=1,
yminorticks=true,
xlabel={$1/\sigma_n^2\ \text{in dB}\rightarrow$},
ylabel={$\text{Gross Symbol Error Rate (GSE)} \rightarrow$},
xmajorgrids,
ymajorgrids,
yminorgrids,
legend entries={$N=20$,$N=10$,$N=5$,$ \text{BPDN}$,$N=2$,$N=1$},
legend style={nodes=right},
legend pos= south west,
legend style={nodes=right}]
\addplot [
color=red,
solid,
line width=1pt,
mark=*,
mark options={solid,fill=white}
]
coordinates{
 (-5,0.199821)(0,0.194289)(5,0.171628)(10,0.116908)(15,0.035993)(20,0.00425352)(25,0.000308169)(30,2.11268e-05) 
};

\addplot [
color=darkgreen,
solid,
line width=1pt,
mark=square*,
mark options={solid,fill=white}
]
coordinates{
 (-5,0.199365)(0,0.198859)(5,0.183283)(10,0.145299)(15,0.0707697)(20,0.008876)(25,0.00057)(30,4.9e-05) 
};

\addplot [
color=blue,
line width=1pt,
solid,
mark=diamond*,
mark options={solid,fill=white}
]
coordinates{
 (-5,0.200948)(0,0.199292)(5,0.193384)(10,0.179949)(15,0.138412)(20,0.0592426)(25,0.00632353)(30,5.88235e-05) 
};

\addplot [
color=blue,
dotted,
line width=1pt,
mark=diamond*,
mark options={solid}
]
coordinates{
 (5,0.407831)(10,0.304715)(15,0.223292)(20,0.147463)(25,0.0903337)(30,0.063621)(35,0.0571667)(40,0.0582487) 
};

\addplot [
color=black,
line width=1pt,
solid,
mark=pentagon*,
mark options={solid,fill=white}
]
coordinates{
 (-5,0.200088)(0,0.202059)(5,0.203235)(10,0.205294)(15,0.198441)(20,0.159265)(25,0.0958824)(30,0.0299412)(35,0.00461765)(40,8.82353e-05) 
};

\addplot [
color=red,
line width=1pt,
solid,
mark=x,
mark options={solid,fill=white}
]
coordinates{
 (-5,0.198786)(0,0.200786)(5,0.210964)(10,0.216)(15,0.2165)(20,0.200071)(25,0.153321)(30,0.0903571)(35,0.0381429)(40,0.0111429) 
};

\end{semilogyaxis}
\end{tikzpicture}
\vspace{-1em}
\caption{Gross Symbol Error Rate parametrized by the length of the spreading sequence $N$. For $N=5$, the dotted curve shows the performance of a Basis Pursuit De-noising for comparison.}\label{fig:GSE}
\end{figure} 
Fig.\ref{fig:GSE} shows the GSE for varying the spreading gain $N$. Most significantly, decreasing from $N=20$ chips down to $N=10$ chips results only in small losses. Overloading the system further, decreases the performance. However, even with $N=1$, communication is still possible at high SNR. One should note that for $N=1$, the system has the dimension $4 \times 20$ due to the channel. Additionally, the BR-CS-MUD performs exhaustive search among all possible hypothesis with an efficient search method. This explains the good performance of our detector. Additionally, the performance of the Basis Pursuit De-noising (BPDN) is shown for a system with a spreading gain of $N=5$ for comparison. Comparing the performance of the BPDN with the BR-CS-MUD shows that the BPDN converges into a error floor even for high SNR. Our detector gains performance from the knowledge that the source vector obeys a constant alphabet with a known a priori distribution. In contrast, the BPDN relaxes the problem to a convex 
optimization problem prior to solving, which results in significant performance losses.

\subsection{Performance of the Activity Detection}
\begin{figure}[ht]
%
%
\begin{tikzpicture}

\begin{axis}[%
scale only axis,
width=6cm,
height=6cm,
xmin=0, xmax=1,
ymin=0, ymax=1,
xlabel={False Active rate},
ylabel={True Active Rate},
xmajorgrids,
ymajorgrids,
legend entries={$\Omega=0.01$,{$\Omega=0.1$},$\Omega=1$,$\Omega=10$,$\Omega=100$},
legend style={nodes=right},
legend pos= south east,
legend style={nodes=right}]
\addplot [
color=red,
solid,
line width=1pt,
mark=*,
mark options={solid,fill=white}
]
coordinates{
 (1,1)(0.997455,0.999003)(0.953428,0.982946)(0.829871,0.925248)(0.601523,0.888549)(0.148299,0.974106)(0.00600916,0.999056)(0.00039986,1) 
};

\addplot [
color=darkgreen,
solid,
line width=1pt,
mark=square*,
mark options={solid,fill=white}
]
coordinates{
 (0.932635,0.96127)(0.823682,0.900297)(0.664496,0.836527)(0.457096,0.837902)(0.130077,0.952476)(0.00938534,0.99673)(0.000537137,0.9999)(6.26433e-05,1) 
};

\addplot [
color=blue,
line width=1pt,
solid,
mark=diamond*,
mark options={solid,fill=white}
]
coordinates{
 (0.0356016,0.105169)(0.0816058,0.334221)(0.0756265,0.553555)(0.0503323,0.805044)(0.0109157,0.963675)(0.00108784,0.996255)(0.000112608,0.999502)(2.50344e-05,0.99995)(0,1) 
};

\addplot [
color=black,
line width=1pt,
solid,
mark=pentagon*,
mark options={solid,fill=white}
]
coordinates{
 (0.00992908,0.0616822)(0.0434468,0.262537)(0.0464824,0.491837)(0.0345161,0.760467)(0.00671598,0.956522)(0.000812429,0.995298)(0.000174961,0.9996)(0,1) 
};

\addplot [
color=red,
line width=1pt,
solid,
mark=x,
mark options={solid,fill=white}
]
coordinates{
 (0,0.00199203)(0.000958773,0.0438931)(0.00687146,0.213836)(0.0123235,0.548329)(0.00614873,0.859614)(0.000574777,0.981271)(0,0.998793) 
};

\addplot [
color=black,
line width=1pt,
solid
]
coordinates{
 (0,1) (1,0)
};

\node[] at (axis description cs:0.3,0.82) {\begin{rotate}{-20}$\Longleftarrow$\end{rotate}};
\node[] at (axis description cs:0.6,0.78) {$\text{SNR Increasing}$};
\node[] at (axis description cs:0.35,0.68) {\begin{rotate}{-45} Liberal\end{rotate}};
\node[] at (axis description cs:0.3,0.63) {\begin{rotate}{-45} Conservative\end{rotate}};

\end{axis}
\end{tikzpicture}
\vspace{-1em}
\caption{Reciever Operating Characteristic for the activity detection for an SNR range of $5\text{dB} -40 \text{dB}$}\label{fig:ROC}
\end{figure}
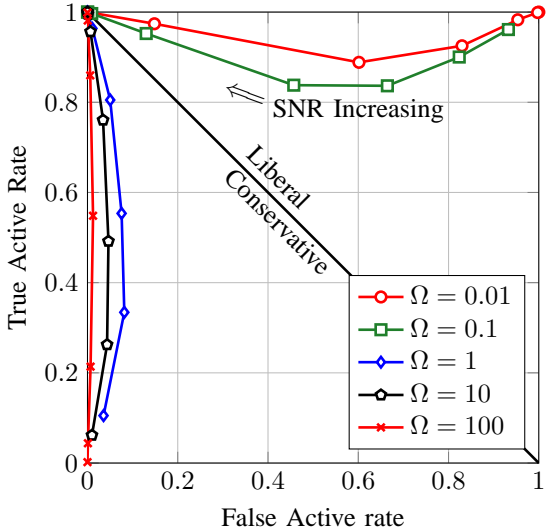

To assess the performance of the activity detection, we employ the so called Receiver Operating Characteristic (ROC) curve~\cite{citeulike:161814}. The ROC evaluates the True Active and the False Active rates of the activity detection which are dependent on the SNR and the choice of $\Omega$. An optimal detector will produce a True Active rate of $100 \%$ and a False Active rate of $0 \%$. Fig.~\ref{fig:ROC} shows the ROC for the setup summarized in Table~\ref{table_sim_par1}. The spreading gain is set to a fixed value of $N=5$ for all nodes and the Bayes factor $\Omega$ is varied to influence the activity detection.
For a Bayes factor $\Omega=1$, the curve starts at the point $(0/0)$ which means that the detector neither correctly detects active nodes nor detects nodes wrongly to be active. Consequently, the detector estimates all nodes to be inactive. For increasing SNR (moving upwards on the curve), this detector increases the True Active rate while still meeting a very low False Active rate. This shows that the detector only detects nodes to be active if the particular probability that the node is active is very high, which can be interpreted as a conservative behavior of the detector. Increasing the Bayes factor to $\Omega=10$ and $\Omega=100$ makes the detector even more conservative and the False Active rate is kept very low at increasing SNR. At very high SNR, the detector reaches the $(1/0)$ point and no errors for the activity detection occur. \\
Decreasing the Bayes factor down to $\Omega=0.1$ and $\Omega=0.01$ promotes the detection of symbols instead of zeros. This setting tunes the detector such that the activity detection decides more likely in favor of activity instead of inactivity. At very low SNR, the curves start at the point $(1/1)$ and the detector estimates all nodes to be active, producing a False Active and True Active rate of $100 \%$. The behavior for increasing SNR is analogous to the previous setting, the False Active rate decreases down to $0 \%$. The True Active Rate also decreases slightly and reaches $100 \%$ for high SNR. Since the detector has the tendency to decide in favor of activity instead of inactivity, we term this a liberal detector. In a practical setup $\Omega$ can be tuned by higher layer applications such that certain False Active of True Active rates are met. Moreover, it might be useful to scale $\Omega$ as a function of the current SNR.

\section{Conclusion}
In this work, we introduced a novel Compressed Sensing Bayes-Risk Detector that minimizes the risk of an erroneous decision with respect to the activity detection. This detector allows reliable communication even in under-determined communication systems. As a degree of freedom the performance of the activity detection is controlled by an additional parameter. This provides a system depended flexible management of the False Active and False Inactive rate and can be adjusted by higher layers such that certain system specific error rates are met. We showed that our detector utilizes the finite alphabet constraint of the modulation alphabet as side information, which allows reliable communication even in highly under-determines systems. With this side information we our detector outperforms a classical Basis Pursuit De-noising approach.

\section*{Acknowledgment}
This work was founded by the German Research 
Foundation (DFG) under grant DE 759/3-1.

\bibliographystyle{IEEEtran}
\bibliography{references}

\begin{thebibliography}{10}
\providecommand{\url}[1]{#1}
\csname url@samestyle\endcsname
\providecommand{\newblock}{\relax}
\providecommand{\bibinfo}[2]{#2}
\providecommand{\BIBentrySTDinterwordspacing}{\spaceskip=0pt\relax}
\providecommand{\BIBentryALTinterwordstretchfactor}{4}
\providecommand{\BIBentryALTinterwordspacing}{\spaceskip=\fontdimen2\font plus
\BIBentryALTinterwordstretchfactor\fontdimen3\font minus
  \fontdimen4\font\relax}
\providecommand{\BIBforeignlanguage}[2]{{%
\expandafter\ifx\csname l@#1\endcsname\relax
\typeout{** WARNING: IEEEtran.bst: No hyphenation pattern has been}%
\typeout{** loaded for the language `#1'. Using the pattern for}%
\typeout{** the default language instead.}%
\else
\language=\csname l@#1\endcsname
\fi
#2}}
\providecommand{\BIBdecl}{\relax}
\BIBdecl

\bibitem{yonia}
Y.~C. Eldar and G.~Kutyniok, \emph{Compressed Sensing: Theory and
  Applications}.\hskip 1em plus 0.5em minus 0.4em\relax Cambridge, U.K.:
  Cambridge Univ. Press, May 2012.

\bibitem{4472240}
E.~Candes and M.~Wakin, ``An introduction to compressive sampling,'' \emph{IEEE
  Signal Processing Magazine}, vol.~25, no.~2, pp. 21 --30, Mar. 2008.

\bibitem{342465}
Y.~Pati, R.~Rezaiifar, and P.~Krishnaprasad, ``Orthogonal matching pursuit:
  recursive function approximation with applications to wavelet
  decomposition,'' in \emph{The Twenty-Seventh Asilomar Conference on Signals,
  Systems and Computers}, nov 1993, pp. 40 -- 44 vol.1.

\bibitem{eps251147}
S.~Chen, S.~A. Billings, and W.~Luo, ``Orthogonal least squares methods and
  their application to non-linear system identification,'' \emph{International
  Journal of Control}, vol.~50, pp. 1873--1896, 1989.

\bibitem{6292852}
R.~F. Fischer and F.~Waeckerle, ``{Peak-to-Average Power Ratio Reduction in
  OFDM via Sparse Signals: Transmitter-Side Tone Reservation vs. Receiver-Side
  Compressed Sensing},'' \emph{17th International OFDM Workshop 2012}, pp. 1
  --8, Aug. 2012.

\bibitem{henn1}
H.~F. Schepker and A.~Dekorsy, ``{Sparse Multi-User Detection for CDMA
  transmission using greedy algorithms},'' in \emph{8th International Symposium
  on Wireless Communication Systems (ISWCS)}, Aachen, Germany, Nov. 2011, pp.
  291 --295.

\bibitem{giannakis}
H.~Zhu and G.~Giannakis, ``Exploiting sparse user activity in multiuser
  detection,'' \emph{IEEE Transactions on Communications}, vol.~59, no.~2, pp.
  454 -- 465, Feb. 2011.

\bibitem{citeulike:161814}
T.~Hastie, R.~Tibshirani, and J.~H. Friedman, \emph{{The Elements of
  Statistical Learning}}, 2nd~ed.\hskip 1em plus 0.5em minus 0.4em\relax
  Springer, Jul. 2011.

\bibitem{VanTDeteRada1971}
H.~L. van Trees, \emph{Detection, estimation and modulation theory}.\hskip 1em
  plus 0.5em minus 0.4em\relax Wiley, 1967.

\bibitem{stephem_kay}
S.~M. Kay, \emph{{Fundamentals of Statistical Signal Processing Detection
  Theory}}, 1st~ed.\hskip 1em plus 0.5em minus 0.4em\relax Prentice Hall, 1993.

\bibitem{Golub1989}
{G. H. Golub and C. F. Van Loan}, \emph{Matrix Computations}, 2nd~ed.\hskip 1em
  plus 0.5em minus 0.4em\relax The Johns Hopkins University Press, 1989.

\bibitem{damen}
M.~O. Damen, H.~E. Gamal, and G.~Caire, ``On maximum-likelihood detection and
  the search for the closest lattice point,'' \emph{IEEE Transactions on
  Information Theory}, vol.~49, no.~10, pp. 2389 -- 2402, oct. 2003.

\bibitem{Verdu}
S.~Verd{\'u}, \emph{Multiuser Detection}.\hskip 1em plus 0.5em minus
  0.4em\relax Cambridge, U.K.: Cambridge Univ. Press, Nov. 1998.

\end{thebibliography}

\end{document}